\documentclass[12pt,a4paper]{article}

\usepackage[utf8]{inputenc}
\usepackage[T1]{fontenc}
\usepackage{lmodern}
\usepackage{amsmath,amssymb,amsthm}
\usepackage{mathrsfs}
\usepackage{tikz-cd}
\usepackage[margin=2.7cm]{geometry}
\usepackage{enumitem}
\usepackage{xcolor}
\usepackage[colorlinks=true,linkcolor=blue!50!black,citecolor=blue!50!black,urlcolor=blue!50!black]{hyperref}

\numberwithin{equation}{section}
\newtheorem{theorem}{Theorem}[section]
\newtheorem{proposition}[theorem]{Proposition}
\newtheorem{lemma}[theorem]{Lemma}
\newtheorem{corollary}[theorem]{Corollary}
\theoremstyle{definition}
\newtheorem{definition}[theorem]{Definition}

\theoremstyle{remark}
\newtheorem{remark}[theorem]{Remark}

\newcommand{\R}{\mathbb{R}}
\newcommand{\E}{\mathbb{E}^3}
\newcommand{\g}{\mathfrak{g}}
\newcommand{\Q}{\mathcal{Q}}
\newcommand{\G}{\mathcal{G}}
\newcommand{\Ad}{\mathrm{Ad}}
\newcommand{\tr}{\operatorname{tr}}

\newcommand{\GL}{GL(3,\R)}
\newcommand{\bpsi}{\bar\psi}
\newcommand{\sh}{^{\sharp}}
\newcommand{\Pio}{\mathsf{P}}   
\newcommand{\Mat}{\mathsf{M}}   
\newcommand{\Esh}{\mathsf{b}}   

\title{On the general Eshelby and Piola entities\\ in material evolution}

\author{
Manuel de Le\'on\\
{\small Instituto de Ciencias Matem\'aticas (CSIC-UAM-UC3M-UCM), Madrid, Spain}\\
{\small Real Academia de Ciencias Exactas, F\'isicas y Naturales de Espa\~na, Madrid, Spain}\\
{\small Great Bay University, Dongguan, Guangdong, China}\\
{\small \texttt{mdeleon@icmat.es}}
\and
Marcelo Epstein\\
{\small Department of Mechanical and Manufacturing Engineering}\\
{\small University of Calgary, Canada}\\
{\small \texttt{mepstein@ucalgary.ca}}
}

\date{}

\begin{document}

\maketitle

\begin{abstract}
In the classical theory of material evolution of simple bodies, the derivative
of the strain-energy density with respect to the local material deformation (the
implant map) is expressible in terms of the Eshelby stress, which in turn is
obtained from the Piola stress and the energy density. We show that this
relation is not an accident of the first-order theory, but a direct consequence
of the differential of a Lie-group action. The configurations of a body of any
grade $k$, with or without microstructure, are represented at each point by
$k$-jets on which the vertical (isotropy) group of local material deformations
acts freely on the right. For an energy density of the form
$\bpsi(q,g)=\rho(g)\,\psi(qg)$, where $\rho$ is a multiplicative density factor,
we prove that the right-trivialized material entity equals minus a
\emph{general Eshelby entity}
$\Esh=-\bpsi\,\vartheta-\sigma_q^{*}\Pio$, living in the dual of the Lie algebra of
the vertical group. Here $\Pio$ is the (generalized) Piola entity, $\sigma_q$ is
the infinitesimal action and $\vartheta$ is the infinitesimal character of~$\rho$.
The formula is explicit for every order, and we work it out for simple bodies
(recovering the classical Eshelby tensor), for second-grade materials and for
bodies with microstructure. We also show that the Eshelby entity annihilates the
Lie algebra of the material symmetry group, and that it is the driving force
conjugate to the right-trivialized rate of material evolution in the
dissipation inequality.
\end{abstract}



\tableofcontents

\section{Introduction}

The energy-momentum tensor introduced by Eshelby \cite{Eshelby51,Eshelby75}
plays a central role in the mechanics of defects, fracture and, more generally,
in the theory of configurational (or material) forces
\cite{Maugin93,Gurtin00,Maugin11}. In the context of material uniformity and
inhomogeneity \cite{Noll67,Wang67,EEbook}, Epstein and Maugin \cite{EM90}
observed that the Eshelby stress arises naturally as the derivative of the
strain-energy density with respect to the \emph{implant} map that places a
fixed archetype at each point of a reference configuration. This observation
underlies the formulation of evolution laws for remodelling, growth and
plasticity in which the Eshelby stress acts as the driving force
\cite{EM00,EEbook}.

In the first-order (simple) theory the computation is elementary: if the energy
per unit reference volume is $W_0(F;P)=J_P^{-1}W(FP)$, then
$\partial W_0/\partial P=-\Esh\,P^{-T}$, where $\Esh=W_0 I-F^{T}T_0$ is the
Eshelby stress and $T_0=\partial W_0/\partial F$ is the Piola stress. When one
tries to repeat the computation for materials of higher grade or for bodies
endowed with a microstructure, the index gymnastics quickly becomes unwieldy,
and the underlying structure is obscured.

The purpose of this paper is to show that the relation between the Piola and the
material (Eshelby) entities is a direct consequence of a very simple fact about
the differential of a group action, namely the formula
\[
\pi_*(v_q,v_g)=R_{g*}(v_q)+\ell_{q*}(v_g)
\]
for the differential of a right action $\pi(q,g)=qg$. The ingredients are:
\begin{itemize}[leftmargin=2em]
\item the manifold $\Q_c$ of $k$-jets at a material point $c$ of the
configurations of the body (ordinary, or with microstructure modelled as a
fibre bundle);
\item the Lie group $\G_c$ of $k$-jets of local material deformations fixing
$c$, which acts freely on $\Q_c$ on the right by composition;
\item a constitutive law of the form $\bpsi(q,g)=\rho(g)\,\psi(qg)$, where
$g\in\G_c$ plays the role of an internal variable describing material evolution
and $\rho$ is a multiplicative density factor.
\end{itemize}
Our main result (Theorem~\ref{thm:main}) states that the material entity
$\Mat=\partial\bpsi/\partial g$, once right-trivialized, equals minus the
\emph{Eshelby entity}
\[
\Esh=-\bpsi\,\vartheta-\sigma_q^{*}\Pio\ \in\g_c^{*},
\]
where $\Pio=\partial\bpsi/\partial q$ is the Piola entity, $\sigma_q:\g_c\to T_q\Q_c$
is the infinitesimal action and $\vartheta=d\rho_e$. Since the infinitesimal action
of a jet is simply composition with the jet of a vector field, the formula is
completely explicit at every order.

The map $\Pio\mapsto\sigma_q^*\Pio$ is the momentum map of the cotangent lift
of the right action of $\G_c$ on $\Q_c$ (Remark~\ref{rem:momentum}). Up to the term
coming from the density factor, the Eshelby entity is therefore minus a momentum
map, in agreement with the classical interpretation of the Eshelby tensor as an
energy-momentum tensor \cite{Eshelby75}. Eshelby stresses and hyperstresses of
second-gradient materials, and evolution laws for second-order implants, have
been obtained by direct computation in \cite{KS07,CM11,CAM12}, and configurational
quantities for micromorphic media are studied in \cite{HKS07}. The present
approach gives a single derivation, valid at every order and for bodies with
microstructure, in which these formulas appear as particular cases of one
identity.

The paper is organized as follows. Section~\ref{sec:review} reviews the basic
differential geometric terminology, with particular emphasis on Lie groups and
their actions. Section~\ref{sec:local} describes the kinematics of bodies with
or without microstructure in terms of jets, and sets up the constitutive
framework. Section~\ref{sec:EP} contains the main results, which are
illustrated in Section~\ref{sec:examples} with simple bodies, second-grade
bodies and bodies with microstructure. Section~\ref{sec:sym} deals with material
symmetries and with the dissipation inequality governing material evolution.

\section{Review and terminology}\label{sec:review}

\subsection{Manifolds}

Somewhat informally, we recall that a topological manifold $M$ of dimension $m$
is a continuous entity (more technically, a topological space) each of whose
points has an open neighbourhood homeomorphic to $\R^m$ (or, equivalently, to
an open set in $\R^m$). A homeomorphism is understood as a continuous bijection
between topological spaces with a continuous inverse. Any such homeomorphism
$\phi$ between an open neighbourhood $U\subset M$ and an open set in $\R^m$
constitutes a pair $(U,\phi)$ known as a \emph{coordinate chart}, since it
assigns $m$ real numbers to each point in $U$. A set of charts whose union is
the whole of $M$ is called an \emph{atlas}, a terminology obviously inspired by
cartography. At the (non-empty) intersection of two charts $(U_1,\phi_1)$ and
$(U_2,\phi_2)$ the \emph{transition function} $\phi_2\circ\phi_1^{-1}$ is a
homeomorphism between two open sets in $\R^m$. A \emph{smooth manifold} is a
topological manifold with an atlas whose transition functions are
diffeomorphisms, namely smooth (infinitely differentiable) homeomorphisms with
smooth inverses.\footnote{The definitions above are lacking in some technical
details, such as second countability and Hausdorff character of the underlying
topological space and maximality of the atlas. See, e.g., \cite{KN63,Lee13}.}

\subsection{Maps}

A map $f:M\to N$ between two smooth manifolds is said to be \emph{smooth} if at
each point $p\in M$ the representation of $f$ in terms of coordinate charts
$U_p$ and $V_{f(p)}$ around $p$ and $f(p)$ is smooth. A \emph{parametrized
curve} (or simply a curve) at a point $p\in M$ is a smooth map
$\gamma:J\to M$, where $J\subset\R$ is an open interval containing the origin,
such that $\gamma(0)=p$. Two parametrized curves $\gamma$ and $\delta$ at $p$ are
\emph{mutually tangent} at $p$ if their coordinate representations in any chart
around $p$ have the same derivative at the origin $0\in\R$. Since tangency at
$p$ is clearly an equivalence relation, we define a \emph{tangent vector} at
$p\in M$ as an equivalence class of tangent curves at $p$. The collection of all
tangent vectors at $p$ constitutes a vector space $T_pM$ called the
\emph{tangent space} at $p\in M$. Its dual space $T_p^*M$ is the
\emph{cotangent space} at $p$.

A curve in the Cartesian product of two manifolds $M_1\times M_2$ consists of a
pair of curves, one in each of the factor manifolds. Consequently, a vector at a
point $(p_1,p_2)\in M_1\times M_2$ is an ordered pair of vectors, one at each of
the factor manifolds. In other words,
$T_{(p_1,p_2)}(M_1\times M_2)=T_{p_1}M_1\oplus T_{p_2}M_2$.

\subsection{Differential of a map}

A curve $\gamma:J\to M$ at $p\in M$ is mapped by $f:M\to N$ to the curve
$f\circ\gamma:J\to N$ at $f(p)\in N$. Since tangency is preserved, the whole
equivalence class constituting a tangent vector at $p$ is mapped by $f$ to a
tangent vector at $f(p)$. By expressing the map $f$ in terms of coordinate
charts it can be shown that linear combinations of vectors at $p$ are mapped
into the corresponding linear combinations of the image vectors. Thus, a linear
map $f_*:T_pM\to T_{f(p)}N$, called the \emph{differential} of $f$ at $p$, is
obtained. The differential of a composition of maps is equal to the composition
of their differentials, which is tantamount to the chain rule of
differentiation (or, when expressed in coordinate charts, to the multiplication
of Jacobian matrices). For a real-valued function $f:M\to\R$ we identify
$T_{f(p)}\R$ with $\R$, so that $f_*$ at $p$ is an element of $T^*_pM$, also
denoted by $df_p$.

For a function $f:M_1\times M_2\to N$, one can introduce the \emph{partial
differentials} $f_{1*}$ and $f_{2*}$ at a point $(p_1,p_2)$ as the respective
differentials of the functions $f_1:M_1\to N$ and $f_2:M_2\to N$ defined as
\begin{equation}\label{eq:partial}
f_1(\cdot)=f(\cdot,p_2),\qquad f_2(\cdot)=f(p_1,\cdot).
\end{equation}
The differential of $f$ evaluated at the vector
$(v_1,v_2)\in T_{(p_1,p_2)}(M_1\times M_2)$ is obtained as
\begin{equation}\label{eq:partial2}
f_*(v_1,v_2)=f_{1*}(v_1)+f_{2*}(v_2),
\end{equation}
where the notation of Equation~\eqref{eq:partial} has been used.

\subsection{Lie groups}

A \emph{Lie group} is a smooth manifold $G$ endowed with a group structure whose
multiplication $G\times G\to G$ and inversion $G\to G$ operations are smooth
functions. If $g,h\in G$ their product will be denoted by $gh$. The inverse of
an element $g\in G$ will be denoted by $g^{-1}$, and the identity element by
$e$.

A \emph{left action} $L$ of the Lie group $G$ on a (smooth) manifold $M$
assigns (smoothly) to each $g\in G$ a diffeomorphism (or transformation)
$L_g:M\to M$ with the property $L_g\circ L_h=L_{gh}$ for every pair $g,h\in G$.
It follows that $L_e$ is the identity transformation of $M$. A similar
definition applies to a \emph{right action} $R$ except that the composition
property now reads $R_g\circ R_h=R_{hg}$. An action is said to be
\emph{effective} if the only element that leaves each point of $M$ fixed is the
group identity, and \emph{free} if the only element that leaves \emph{some}
point of $M$ fixed is the group identity. The Lie group structure induces
naturally a free left action and a free right action on itself, called the
\emph{left} and the \emph{right translations}, denoted by $L$ and $R$, and
defined, respectively, by
\[
L_g(h)=gh\qquad\text{and}\qquad R_g(h)=hg.
\]

To obtain an expression for the differential of the group multiplication
$\pi:G\times G\to G$, $\pi(g,h)=gh$, we follow the notation \eqref{eq:partial}
and evaluate
\begin{equation}\label{eq:pi12}
\pi_1(g)=\pi(g,h)=gh=R_h(g),\qquad \pi_2(h)=\pi(g,h)=gh=L_g(h),
\end{equation}
where we have carefully distinguished between the variables in each case.
Equation~\eqref{eq:partial2} yields
\begin{equation}\label{eq:dpi}
\pi_*(v_g,v_h)=R_{h*}(v_g)+L_{g*}(v_h),
\end{equation}
where $v_g$ and $v_h$ are vectors tangent to $G$ at $g$ and $h$,
respectively.

Let
\begin{equation}\label{eq:psi}
\psi:G\to\R
\end{equation}
be a real-valued function on a Lie group $G$. We can define an associated
function $\bpsi:G\times G\to\R$ by setting
\begin{equation}\label{eq:psibar}
\bpsi(g,h)=\psi(gh)=\psi\circ\pi(g,h).
\end{equation}
By the chain rule
\begin{equation}\label{eq:dpsibar}
\bpsi_*(v_g,v_h)=\psi_*\big|_{gh}\circ\pi_*(v_g,v_h)
=\psi_*\big|_{gh}\circ\big(R_{h*}(v_g)+L_{g*}(v_h)\big).
\end{equation}

\subsection{Lie algebra, trivializations and the adjoint representation}

The \emph{Lie algebra} $\g$ of $G$ is the tangent space $T_eG$ endowed with
the bracket of left-invariant vector fields. The \emph{exponential map}
$\exp:\g\to G$ assigns to $A\in\g$ the value at $t=1$ of the one-parameter
subgroup $t\mapsto\exp(tA)$ with initial velocity $A$. The \emph{conjugation}
$I_g(h)=ghg^{-1}=L_g\circ R_{g^{-1}}(h)$ fixes $e$; its differential at $e$ is
the \emph{adjoint representation}
\begin{equation}\label{eq:Ad}
\Ad_g=(I_g)_{*e}=L_{g*}\circ R_{g^{-1}*}:\g\to\g,
\qquad \Ad_g\Ad_h=\Ad_{gh}.
\end{equation}
The \emph{coadjoint} map $\Ad^*_g:\g^*\to\g^*$ is the transpose of $\Ad_g$,
namely $(\Ad^*_g\beta)(A)=\beta(\Ad_gA)$.

Since $L_{g*}$ and $R_{g*}$ are linear isomorphisms $\g\to T_gG$, every
covector $\alpha\in T^*_gG$ can be \emph{trivialized} in two ways:
\begin{equation}\label{eq:triv}
\alpha^{L}=\alpha\circ L_{g*}\in\g^*,\qquad
\alpha^{R}=\alpha\circ R_{g*}\in\g^*,\qquad
\alpha^{L}=\Ad^*_g\,\alpha^{R}.
\end{equation}
The last identity follows from $L_{g*}=R_{g*}\circ\Ad_g$, which is a
consequence of \eqref{eq:Ad}. For a matrix group, $L_{g*}A=gA$,
$R_{g*}A=Ag$ and $\Ad_gA=gAg^{-1}$.

A smooth map $\rho:G\to\R^+$ is a (positive) \emph{character} if
$\rho(gh)=\rho(g)\rho(h)$. Its \emph{infinitesimal character} is
$\vartheta=d\rho_e\in\g^*$. Differentiating $\rho(g\exp(tA))=\rho(g)\rho(\exp(tA))$
and $\rho(\exp(tA)g)=\rho(\exp(tA))\rho(g)$ at $t=0$ we obtain
\begin{equation}\label{eq:char}
d\rho_g\circ L_{g*}=d\rho_g\circ R_{g*}=\rho(g)\,\vartheta,
\end{equation}
so that, by \eqref{eq:triv}, $\vartheta\circ\Ad_g=\vartheta$ for all $g\in G$. The
typical example is $\rho(g)=|\det g|^{-1}$ on $GL(n,\R)$, for which
$\vartheta(A)=-\tr A$.

\subsection{Actions on manifolds and fundamental vector fields}\label{sec:actions}

Let $G$ act smoothly on the right on a manifold $Q$, and denote the action by
\[
\pi:Q\times G\to Q,\qquad \pi(q,g)=qg.
\]
Besides the transformations $R_g:Q\to Q$, $R_g(q)=qg$, we have for each $q\in Q$
the \emph{orbit map}
\[
\ell_q:G\to Q,\qquad \ell_q(g)=qg.
\]
Exactly as in \eqref{eq:pi12}--\eqref{eq:dpi}, the differential of the action
reads
\begin{equation}\label{eq:dpiQ}
\pi_*(v_q,v_g)=R_{g*}(v_q)+\ell_{q*}(v_g),\qquad v_q\in T_qQ,\ v_g\in T_gG.
\end{equation}
The \emph{infinitesimal action} at $q$ is the linear map
\begin{equation}\label{eq:sigma}
\sigma_q=\ell_{q*}\big|_e:\g\to T_qQ,\qquad
\sigma_q(A)=A\sh(q)=\frac{d}{dt}\Big|_{t=0}q\exp(tA),
\end{equation}
and $A\sh$ is called the \emph{fundamental vector field} generated by $A$. If
the action is free, $\sigma_q$ is injective for every $q$. Two elementary
identities will be crucial in what follows. First, since
$\ell_q\circ L_g=\ell_{qg}$,
\begin{equation}\label{eq:ellL}
\ell_{q*}\circ L_{g*}=\sigma_{qg}.
\end{equation}
Second, since $q\exp(tA)\,g=qg\,\big(g^{-1}\exp(tA)g\big)=qg\exp(t\Ad_{g^{-1}}A)$,
\begin{equation}\label{eq:Requiv}
R_{g*}\big(\sigma_q(A)\big)=\sigma_{qg}\big(\Ad_{g^{-1}}A\big).
\end{equation}
When $Q=G$ with the right translation, $\ell_q=L_q$ and \eqref{eq:dpiQ} reduces
to \eqref{eq:dpi}.

\section{Groups of local deformations}\label{sec:local}

\subsection{Bodies and microstructure}

In continuum mechanics, and most especially in the mechanics of solid media,
one of the foundational elements of the discipline is the precise geometric
formulation of the notions of material bodies and their deformations in space
\cite{Noll67,Truesdell,Epstein10}. At its most basic level, which is sufficient
for many engineering applications, a material body is a connected
3-dimensional manifold $\mathcal B$ that can be covered with a single chart. A
\emph{configuration} of $\mathcal B$ is a map of $\mathcal B$ into the Euclidean
space $\E$ of classical physics. This map $\kappa:\mathcal B\to\E$ is at least
continuous and one-to-one and is often assumed to have some degree of
differentiability. In the extreme case, one can assume that $\kappa$ is a
smooth diffeomorphism onto its image $\kappa(\mathcal B)\subset\E$.

In other important engineering applications the body may include some extra
internal structure, such as in the case of granular materials in which the
grains constitute a so-called \emph{microstructure}, embedded into a
substratum, or \emph{macrostructure}, with which it interacts. Each grain is a
continuous medium in its own right, but its deformations are described by a
finite number of kinematic parameters. The mathematical representation of a
body with microstructure is nicely encapsulated in the theory of fibre bundles
\cite{Capriz89,EdL98}, consisting of a base manifold $\mathcal B$ at each of
whose points a fibre is assigned. The base manifold is an ordinary body
representing the substratum, while the nature of the fibres depends on the
application at hand. A deformation of a body with microstructure can be
described as a fibre-bundle morphism, a concept to be briefly reviewed below.

\subsection{Brief review of fibre bundles}

The Cartesian product $\mathcal C=\mathcal B\times\mathcal F$ of two manifolds,
$\mathcal B$ and $\mathcal F$, gives rise to a new manifold $\mathcal C$ called a
\emph{product fibre bundle} with base $\mathcal B$ and typical fibre
$\mathcal F$. A bundle chart is the Cartesian product of two charts, one in each
of the factors. Moreover, since an element $c$ in $\mathcal C$ consists of a pair
$c=(p,r)$, with $p\in\mathcal B$ and $r\in\mathcal F$, a product bundle is
automatically endowed with two projection maps,
$\pi_1:\mathcal C\to\mathcal B$ and $\pi_2:\mathcal C\to\mathcal F$, defined
respectively by $\pi_1(p,r)=p$ and $\pi_2(p,r)=r$. For each $p\in\mathcal B$,
the inverse image $\pi_1^{-1}(p)$ is called the \emph{fibre} over $p$, hence the
terminology of fibre bundle, whereby $\mathcal C$ can be visualized as a
collection of fibres numbered and kept together, as it were, by $\mathcal B$.
It follows trivially from the definition that the projection maps are both
surjective.

In a product bundle there is a definite privileged one-to-one correspondence
between the fibres. Indeed, two points, $(p_1,r_1)$ and $(p_2,r_2)$,
respectively in the fibres $\pi_1^{-1}(p_1)$ and $\pi_1^{-1}(p_2)$, can be
declared to correspond to each other if $r_1=r_2$. For example, if we regard a
2-cylinder in $\R^3$ with Cartesian coordinates $x,y,z$ as the product
$S^1\times\R$, where $S^1$ is a circle in the plane $z=0$, the canonically
corresponding points in the vertical generators are located at circles parallel
to the base, at constant values of $z$. We want to create a more general notion
of fibre bundle in which the second projection is eliminated, thus abolishing
the privileged correspondence between points in different fibres. A good, but by
no means the only, motivation for this more liberal concept is the very notion
of Galilean space-time, where the base manifold $\R$ is time and the typical
fibre is $\E$. The existence of the first projection indicates that
simultaneity of events is universally accepted, whereas spatial locations at
different times are literally in the eyes of the beholder, since different
observers are in a state of relative uniform motion.

A \emph{smooth fibre bundle} $(\mathcal C,\mathcal B,\mathcal F,\pi)$ with base
manifold $\mathcal B$ and typical fibre $\mathcal F$ is a manifold $\mathcal C$
and a smooth surjective \emph{bundle projection} map
$\pi:\mathcal C\to\mathcal B$ such that $\mathcal B$ can be covered with an atlas
$(U_\alpha,\phi_\alpha)$ enjoying a \emph{trivialization} property.
Intuitively, we want to think of $\mathcal C$ as covered with the union of all
the sets $V_\alpha=\pi^{-1}(U_\alpha)$, but we also demand that each $V_\alpha$
be diffeomorphic to the product $U_\alpha\times\mathcal F$ in such a way that
fibres in $V_\alpha$ are entirely mapped into fibres in
$U_\alpha\times\mathcal F$. In simpler words, we are conceiving of a fibre
bundle as a trivializable entity, namely, as a bunch of atlas-dependent product
bundles. The transition functions between intersecting charts in the atlas of
the base manifold are naturally diffeomorphisms. The transition functions on the
corresponding fibres are assumed to belong to a preassigned \emph{structure
group} $G$ acting effectively on the left on $\mathcal F$. These transition
functions must also vary smoothly from fibre to neighbouring fibre
\cite{Steenrod51,KN63}. A fibre-preserving smooth map between fibre bundles is
known as a \emph{fibre-bundle morphism}.\footnote{For simplicity, we are
assuming that the structure groups of both bundles are the same.}

\subsection{Configurations, deformations, and their jets}

For an ordinary body $\mathcal B$ we have already defined the notion of
(smooth) configuration as a smooth map $\kappa:\mathcal B\to\E$ which is a
diffeomorphism onto its image. The \emph{deformation} between two
configurations $\kappa_1$ and $\kappa_2$ is defined as the diffeomorphism
$\chi=\kappa_2\circ\kappa_1^{-1}:\kappa_1(\mathcal B)\to\kappa_2(\mathcal B)$.
Two deformations, $\chi_1$ and $\chi_2$, can be composed to yield a deformation
$\chi_2\circ\chi_1$ whenever the target of $\chi_1$ coincides with the source of
$\chi_2$. Technically, therefore, the collection of all the smooth deformations
of a body constitutes a \emph{groupoid} \cite{Mackenzie05}.

In the case of a body with microstructure, modelled as a fibre bundle
$(\mathcal C,\mathcal B,\mathcal F,\pi)$, an appropriate physical space where
configurations can manifest themselves is the product bundle
$\mathcal S=\E\times\mathcal F$. A configuration of a body with microstructure
is defined as a fibre-bundle morphism $K$ from the body bundle to
$\mathcal S$. It subsumes an ordinary configuration $\kappa$ of the substratum
$\mathcal B$. The commutative diagram \eqref{eq:diag1} is a useful
representation of a configuration of a body with microstructure. Similarly,
deformations can be defined as compositions of configurations in the category of
fibre bundles.
\begin{equation}\label{eq:diag1}
\begin{tikzcd}
\mathcal C \arrow[r,"K"] \arrow[d,"\pi"'] & \mathcal S \arrow[d,"\pi_1"]\\
\mathcal B \arrow[r,"\kappa"'] & \E
\end{tikzcd}
\end{equation}

A smooth map $f:M\to N$ between two smooth manifolds, $M$ and $N$, regardless of
any additional structure that these manifolds may convey, is ultimately
representable in some chosen coordinate charts at $p$ and $f(p)$ by smooth
real-valued functions of real variables. We say that two such maps, $f$ and
$g$, between $M$ and $N$ have the same \emph{$k$-jet} at the point $p\in M$ if
the values of $f$ and $g$ and of all their partial derivatives up to and
including the order $k$ in some (and therefore in any) charts coincide at $p$.
We may visualize this property by stating that the two functions have a
\emph{contact of order $k$} at $p$. Having the same $k$-jet at a point is
obviously an equivalence relation, which leads us to identify a $k$-jet as an
equivalence class of maps under this relation
\cite{KMS93,Saunders89}. The $k$-jet of the map $f:M\to N$ at $p\in M$ is
denoted by $j^k_pf$. The collection of all $k$-jets at $p$ of maps from $M$ to
$N$ is denoted by $J^k_p(M,N)$; it is a finite-dimensional manifold. In the case
of configurations and deformations of a body, $k$-jets are known as \emph{local
configurations} and \emph{local deformations} of order $k$.

Moreover, $k$-jets can be composed by identifying the composition of two
$k$-jets (of, say, deformations) as the $k$-jet of the composition of any
representatives of the corresponding $k$-jets:
\[
j^k_{f(p)}h\circ j^k_pf=j^k_p(h\circ f).
\]
Under the operation of composition, invertible $k$-jets form a groupoid. The
so-called \emph{vertical groups} (or isotropy groups) of this groupoid consist of
the $k$-jets of deformations that map a point to itself. For an ordinary body
$\mathcal B$ with $k=1$ these groups are isomorphic to the general linear group
$\GL$. For arbitrary $k$ they are isomorphic to the jet group $G^k_3$ of
$k$-jets at $0\in\R^3$ of local diffeomorphisms of $\R^3$ fixing $0$.

We now fix the objects that will be used in the remainder of the paper. Let
$\mathcal C_0$ be a reference configuration of the body (ordinary, in which case
$\mathcal C_0=\kappa_0(\mathcal B)\subset\E$, or with microstructure, in which
case $\mathcal C_0=K_0(\mathcal C)\subset\mathcal S$), and let $c\in\mathcal C_0$.

\begin{definition}\label{def:QG}
\begin{enumerate}[label=(\roman*)]
\item The \emph{manifold of local deformations of order $k$ at $c$}, denoted
$\Q_c$, is the set of $k$-jets $j^k_c\chi$ at $c$ of (local) deformations
$\chi$ of $\mathcal C_0$ into the physical space, namely, local diffeomorphisms
into $\E$ in the ordinary case, or local fibre-bundle morphisms into
$\mathcal S$ which are diffeomorphisms onto their images in the case of a body
with microstructure. It is an open submanifold of the corresponding jet manifold
$J^k_c(\mathcal C_0,\cdot)$, or of its fibred analogue.
\item The \emph{group of local material deformations of order $k$ at $c$},
denoted $\G_c$, is the vertical group at $c$ of the groupoid of invertible
$k$-jets of local diffeomorphisms (respectively, local fibre-bundle
automorphisms) of $\mathcal C_0$. Its elements are the $k$-jets $j^k_c\zeta$ of
local deformations $\zeta$ of the reference configuration into itself with
$\zeta(c)=c$. Its Lie algebra $\g_c$ consists of the $k$-jets at $c$ of
(projectable, in the bundle case) vector fields on $\mathcal C_0$ vanishing at
$c$.
\end{enumerate}
\end{definition}

The group $\G_c$ acts on $\Q_c$ on the right by composition,
\begin{equation}\label{eq:actionjets}
\Q_c\times\G_c\to\Q_c,\qquad \big(j^k_c\chi,\,j^k_c\zeta\big)\mapsto
j^k_c\chi\circ j^k_c\zeta=j^k_c(\chi\circ\zeta),
\end{equation}
and this action is free, since $q\,g=q$ implies $g=q^{-1}q$ is the identity
jet. Notice that this action does not change the spatial point $\chi(c)$; in
other words, the orbits of $\G_c$ are the fibres of the target map
$\Q_c\to\E$ (respectively $\Q_c\to\mathcal S$), on each of which $\G_c$ acts
simply transitively.

\begin{lemma}\label{lem:sigmajets}
Let $q=j^k_c\chi\in\Q_c$ and let $A=j^k_c\xi\in\g_c$, where $\xi$ is a vector
field on $\mathcal C_0$ with $\xi(c)=0$. Then
\begin{equation}\label{eq:sigmajet}
\sigma_q(A)=j^k_c\big(T\chi\circ\xi\big),
\end{equation}
where the tangent space $T_q\Q_c$ is identified with the space of $k$-jets at $c$
of vector fields along $\chi$ (i.e., of sections of $\chi^*T\E$, respectively
$\chi^*T\mathcal S$). The image of $\sigma_q$ consists of the jets whose
$0$-th order part vanishes at $c$.
\end{lemma}

\begin{proof}
If $\phi_t$ denotes the local flow of $\xi$, then $\phi_t(c)=c$ and
$\exp(tA)=j^k_c\phi_t$. Hence $q\exp(tA)=j^k_c(\chi\circ\phi_t)$ and
\[
\frac{d}{dt}\Big|_{t=0}j^k_c(\chi\circ\phi_t)
=j^k_c\Big(\frac{d}{dt}\Big|_{t=0}\chi\circ\phi_t\Big)
=j^k_c(T\chi\circ\xi),
\]
since differentiation with respect to the parameter $t$ commutes with the
partial derivatives with respect to the coordinates of $\mathcal C_0$ that
define the jet.
\end{proof}

In other words, the infinitesimal action is nothing but the composition of the
jet of the deformation with the jet of the infinitesimal material deformation,
and it can be computed at any order by the ordinary rules of differentiation of
composite functions (see Section~\ref{sec:examples}).

\subsection{Constitutive considerations}

The study and modelling of the enormous variety of material responses observed
in nature and in artificially manufactured materials is known as
\emph{constitutive theory}. Materials that, when studied at a constant
temperature and far from the interference of non-mechanical fields, behave in a
thermodynamically reversible (non-dissipative) way can be characterized by a
single scalar function known as the \emph{strain energy}, which represents the
energy stored in the body and recoverable as mechanical work. This work depends
only on the present state of deformation of the body, regardless of the process
by which this state has been achieved. These materials are also said to be
\emph{hyperelastic}. This designation is often applied only to \emph{local
materials}, or \emph{materials of grade $k$}, in which the strain energy is
given by a density (per unit volume, or per unit mass, say) which depends at each
point only on the $k$-jet of the deformation at that point. To express this
$k$-jet it is customary to adopt a fixed reference configuration $K_0$
relative to which all other configurations can be obtained via a deformation. A
local material of grade 1 is also known as a \emph{simple} hyperelastic
material. It is worth remarking that if the material carries some internal
structure, in other words if the body is modelled as a fibre bundle over an
ordinary body $\mathcal B$, the independent variable of the constitutive
equation is the $k$-jet of the bundle deformation $X$, which also includes the
$k$-jet of the base deformation $\chi$, as can be understood from the
commutative diagram \eqref{eq:diag2}.
\begin{equation}\label{eq:diag2}
\begin{tikzcd}
K_0(\mathcal C) \arrow[r,"X"] \arrow[d,"\pi"'] & \mathcal S \arrow[d,"\pi_1"]\\
\kappa_0(\mathcal B) \arrow[r,"\chi"'] & \E
\end{tikzcd}
\end{equation}

The reference configuration inherits the metric structure of $\E$, thereby
inducing the standard Cartesian volume form in $\kappa_0(\mathcal B)$. Denoting
by $W^c_0:\Q_c\to\R$ the strain-energy function per unit volume of
$\kappa_0(\mathcal B)$ at a fixed $c\in K_0(\mathcal C)$, we define the
\emph{Piola stress entity} at $c$ as the differential of $W^c_0$, that is,
\begin{equation}\label{eq:Piola0}
\Pio^c_0=\big(W^c_0\big)_*\ \in T^*_q\Q_c,\qquad q=j^k_cX.
\end{equation}
The Piola stress entity is of course dependent on the reference configuration
adopted. With the notation of \eqref{eq:Wbar2} below, $W^c_0=\bpsi(\cdot,g)$
for the current local material deformation $g$, and $\Pio^c_0$ is the Piola entity
of Definition~\ref{def:entities}. Since $T_q\Q_c$ consists of $k$-jets of vector fields along $X$
(Lemma~\ref{lem:sigmajets}), the Piola entity is a linear functional on
$k$-jets of virtual displacements: for $k=1$ it comprises the usual Piola
stress, and for $k\ge2$ it includes the hyperstresses of the corresponding
higher-grade theory.

Processes of material evolution, such as remodelling in biological materials,
cannot be considered as thermodynamically reversible in general. Nevertheless,
some of these processes can be satisfactorily described by an energy function
depending not only on the $k$-jet of the deformation but also on a list of
internal variables with a relevant physical meaning, such as the presence of
chemical reactions. Among all the possible choices of such internal variables,
there is a natural choice represented by the right action of a $k$-jet that may
evolve in time. The physical interpretation of this particular choice can be
loosely described as the only choice that preserves the `chemical' identity of
the material. Imagine, as an elementary example, an elastic spring characterized
by a stiffness constant and a resting length. Of these two parameters, the
stiffness is measurable as the coefficient of proportionality between a
variation of the instantaneous spring length and the force necessary to produce
it. The resting length, on the other hand, is a reference length that would be
instantaneously acquired by the spring if it were unloaded. This reference
length could be considered as an internal variable. Similarly, in processes of
bone remodelling, whether natural or artificially enhanced (such as
orthodontics), the realignment of trabeculae, while clearly affecting the
material response, does not necessarily affect the chemical composition of the
bone matrix. The same remarks can be applied to processes of dislocation of
crystalline solids \cite{EEbook,EM00}.

Accordingly, we fix an \emph{archetypal} response $\psi:\Q_c\to\R$, namely the
energy per unit volume of a local archetype, and we assume that the
evolving material is characterized by an energy per unit volume of the reference
configuration of the form $\bpsi:\Q_c\times\G_c\to\R$ given by
\begin{equation}\label{eq:Wbar}
\bpsi\big(j^k_cX,\,j^k_cX_m\big)=\rho\big(j^k_cX_m\big)\,
\psi\big(j^k_cX\circ j^k_cX_m\big),
\end{equation}
where $X_m$ represents a local material deformation (the \emph{implant} of the
archetype at $c$), and $\rho:\G_c\to\R^+$ is a positive character accounting
for the change of the unit of volume. The natural choice is
\begin{equation}\label{eq:rho}
\rho(g)=\big|\det\,(\text{first-order base part of }g)\big|^{-1},
\end{equation}
which is a character because the projection of a $k$-jet onto the $1$-jet of
the underlying base map is a group homomorphism $\G_c\to GL(T_{\pi(c)}\kappa_0(\mathcal
B))\cong\GL$. The case $\rho\equiv1$ corresponds to energies per unit volume of
the archetype rather than of the reference configuration. In what follows we
abbreviate $q=j^k_cX$ and $g=j^k_cX_m$, so that \eqref{eq:Wbar} reads simply
\begin{equation}\label{eq:Wbar2}
\bpsi(q,g)=\rho(g)\,\psi(qg)=\rho(g)\,\psi\circ\pi(q,g).
\end{equation}

\section{The Piola and the Eshelby entities}\label{sec:EP}

\subsection{The Eshelby--Piola entity}

Applying the formalism of Equation~\eqref{eq:dpsibar}, now with the action
formula \eqref{eq:dpiQ} instead of \eqref{eq:dpi}, we obtain what one may call
the \emph{Eshelby--Piola entity}, namely the full differential of $\bpsi$:
\begin{equation}\label{eq:EP}
\bpsi_*(V_q,V_g)=d\rho_g(V_g)\,\psi(qg)
+\rho(g)\,\psi_*\big|_{qg}\circ\big(R_{g*}(V_q)+\ell_{q*}(V_g)\big),
\end{equation}
for all $V_q\in T_q\Q_c$, $V_g\in T_g\G_c$. We recognize here the sum of a
Piola entity and a material entity:

\begin{definition}\label{def:entities}
Let $(q,g)\in\Q_c\times\G_c$.
\begin{enumerate}[label=(\roman*)]
\item The \emph{Piola entity} is the partial differential of $\bpsi$ with
respect to the deformation:
\begin{equation}\label{eq:Pio}
\Pio=\bpsi_{1*}=\rho(g)\,\psi_*\big|_{qg}\circ R_{g*}\ \in T_q^*\Q_c .
\end{equation}
\item The \emph{material entity} is the partial differential of $\bpsi$ with
respect to the local material deformation:
\begin{equation}\label{eq:Mat}
\Mat=\bpsi_{2*}=\psi(qg)\,d\rho_g+\rho(g)\,\psi_*\big|_{qg}\circ\ell_{q*}\
\in T_g^*\G_c .
\end{equation}
\item The \emph{Eshelby entity} is the element of the dual $\g_c^*$ of the Lie
algebra of local material deformations defined by
\begin{equation}\label{eq:Esh}
\Esh=-\bpsi(q,g)\,\vartheta-\sigma_q^*\Pio,
\qquad\text{i.e.}\qquad
\Esh(A)=-\bpsi(q,g)\,\vartheta(A)-\Pio\big(\sigma_q(A)\big),\quad A\in\g_c,
\end{equation}
where $\vartheta=d\rho_e$ is the infinitesimal character of $\rho$.
\end{enumerate}
\end{definition}

The Eshelby entity is thus defined exclusively in terms of the energy density
and the Piola entity, both evaluated at the actual state $(q,g)$, and of the
infinitesimal action $\sigma_q$, which by Lemma~\ref{lem:sigmajets} is simply the
composition of $q$ with $k$-jets of vector fields vanishing at $c$. For
$\rho\equiv1$ it reduces to $-\sigma_q^*\Pio$, which, as we shall see, is the
generalization of (minus) the Mandel stress.

\subsection{Main result}

The question raised by Equation~\eqref{eq:EP} is whether the material entity
can be read off from the Piola entity, by extracting the common factor
$\psi_*|_{qg}$. The following result shows that this is indeed the case for any
order $k$ and for bodies with or without microstructure.

\begin{theorem}\label{thm:main}
Let $\bpsi(q,g)=\rho(g)\psi(qg)$ as in \eqref{eq:Wbar2}. Then the right
trivialization of the material entity is minus the Eshelby entity:
\begin{equation}\label{eq:main}
\Mat\circ R_{g*}=-\Esh .
\end{equation}
Equivalently, the left trivialization of the material entity is
$\Mat\circ L_{g*}=-\Ad^*_g\Esh$, and, for every $V_g\in T_g\G_c$,
\begin{equation}\label{eq:main2}
\Mat(V_g)=-\Esh\big(R_{g^{-1}*}V_g\big)
=\bpsi(q,g)\,\vartheta\big(R_{g^{-1}*}V_g\big)
+\Pio\Big(R_{g^{-1}*}\,\ell_{q*}V_g\Big).
\end{equation}
\end{theorem}

\begin{proof}
Let $A\in\g_c$. We first compute the left trivialization of $\Mat$. By
\eqref{eq:char}, $d\rho_g(L_{g*}A)=\rho(g)\vartheta(A)$, so the first term of
\eqref{eq:Mat} contributes $\rho(g)\psi(qg)\vartheta(A)=\bpsi(q,g)\vartheta(A)$. For the
second term, \eqref{eq:ellL} and \eqref{eq:Requiv} yield
\[
\ell_{q*}(L_{g*}A)=\sigma_{qg}(A)=R_{g*}\big(\sigma_q(\Ad_gA)\big).
\]
Therefore, by \eqref{eq:Pio},
\[
\rho(g)\,\psi_*\big|_{qg}\big(\ell_{q*}L_{g*}A\big)
=\rho(g)\,\psi_*\big|_{qg}\Big(R_{g*}\big(\sigma_q(\Ad_gA)\big)\Big)
=\Pio\big(\sigma_q(\Ad_gA)\big).
\]
Adding both contributions and using the $\Ad$-invariance $\vartheta(A)=\vartheta(\Ad_gA)$,
\[
\Mat(L_{g*}A)=\bpsi\,\vartheta(\Ad_gA)+\Pio\big(\sigma_q(\Ad_gA)\big)
=-\Esh(\Ad_gA)=-\big(\Ad^*_g\Esh\big)(A).
\]
This proves the statement about the left trivialization. By \eqref{eq:triv},
$\Mat\circ R_{g*}=\Ad^*_{g^{-1}}(\Mat\circ L_{g*})=-\Ad^*_{g^{-1}}\Ad^*_g\Esh=-\Esh$,
which is \eqref{eq:main}. Finally, \eqref{eq:main2} follows by writing
$V_g=R_{g*}(R_{g^{-1}*}V_g)$, and using that
$\sigma_q\circ R_{g^{-1}*}=R_{g^{-1}*}\circ\ell_{q*}$ on $T_g\G_c$, which is the
differential at $g$ of the identity $R_{g^{-1}}\circ\ell_q=\ell_q\circ R_{g^{-1}}$
(on the left-hand side $R_{g^{-1}}$ denotes the action on $\Q_c$, and on the
right-hand side the right translation of $\G_c$).
\end{proof}

\begin{remark}
Formula \eqref{eq:main2} answers the question of extracting the common factor
in \eqref{eq:EP} in the most direct way: if $V_g$ is a variation of the local
material deformation, the vector
$R_{g^{-1}*}\ell_{q*}V_g\in T_q\Q_c$ is a variation of the local deformation which
produces exactly the same variation of the composite $qg$ as $V_g$ does (indeed,
$R_{g*}R_{g^{-1}*}\ell_{q*}V_g=\ell_{q*}V_g$). In the group case
$\Q_c=\G_c=G$ this is the map $R_{h^{-1}*}\circ L_{g*}$ obtained by comparing
the two terms of \eqref{eq:dpsibar}.
\end{remark}

\begin{remark}
The Eshelby entity only involves the restriction of $\Pio$ to the image of
$\sigma_q$, namely to the jets that are vertical with respect to the target map,
that is, those that do not move the spatial point $X(c)$. In particular, a
possible dependence of $\psi$ on the position in space (the $0$-th order part of
the jet), which would give rise to a body-force-like component of $\Pio$, never
contributes to the Eshelby entity. Since the action is free, $\sigma_q$ is
injective, so that $\sigma_q^*:T^*_q\Q_c\to\g_c^*$ is onto.
\end{remark}

\begin{remark}\label{rem:momentum}
The right action of $\G_c$ on $\Q_c$ lifts to $T^*\Q_c$ by
$\alpha_q\mapsto\alpha_q\circ R_{g^{-1}*}\in T^*_{qg}\Q_c$, and the momentum map
of this lifted action is
\[
J:T^*\Q_c\to\g_c^*,\qquad J(\alpha_q)=\sigma_q^*\alpha_q .
\]
By \eqref{eq:Requiv}, $J(\alpha_q\circ R_{g^{-1}*})=\Ad^*_gJ(\alpha_q)$. With this
notation, Definition~\ref{def:entities} and Theorem~\ref{thm:main} read
\[
\Esh=-\bpsi\,\vartheta-J(\Pio),\qquad \Mat\circ R_{g*}=\bpsi\,\vartheta+J(\Pio),
\]
and Corollary~\ref{cor:transf} below is a consequence of the
$\Ad^*$-equivariance of $J$, since $\Pio=\rho(g)\,\psi_*|_{qg}\circ R_{g*}$.
\end{remark}

\subsection{Change of implant}

The Eshelby entity of the evolving material can be expressed in terms of the
Eshelby entity of the archetype, defined as
\[
\Esh_\psi(q):=\Esh(q,e)=-\psi(q)\,\vartheta-\sigma_q^*(\psi_*|_q)\in\g_c^*,
\]
where we have used $\rho(e)=1$.

\begin{corollary}\label{cor:transf}
For every $(q,g)\in\Q_c\times\G_c$,
\begin{equation}\label{eq:transf}
\Esh(q,g)=\rho(g)\,\Ad^*_{g^{-1}}\,\Esh_\psi(qg).
\end{equation}
\end{corollary}

\begin{proof}
By \eqref{eq:Pio} and \eqref{eq:Requiv},
$\Pio(\sigma_q(A))=\rho(g)\psi_*|_{qg}(\sigma_{qg}(\Ad_{g^{-1}}A))$. Moreover,
$\bpsi\vartheta(A)=\rho(g)\psi(qg)\vartheta(\Ad_{g^{-1}}A)$. Hence
$\Esh(q,g)(A)=\rho(g)\,\Esh_\psi(qg)(\Ad_{g^{-1}}A)$.
\end{proof}

\section{Examples}\label{sec:examples}

In this section we use Cartesian coordinates $X^I$ in the reference
configuration and $x^i$ in space, and the summation convention. For matrices we
write $A:B=\tr(A^TB)$.

\subsection{Simple bodies: the classical Eshelby stress}

Let $k=1$ and let the body be ordinary. A $1$-jet $q\in\Q_c$ is a pair
$(x,F)$, where $x=\chi(c)$ and $F=D\chi(c)\in\GL$ is the deformation gradient,
while $\G_c\cong\GL$ (we restrict ourselves to the component
$\det P>0$). The action is $(x,F)P=(x,FP)$, and, by \eqref{eq:sigmajet}, the
infinitesimal action of $A\in\mathfrak{gl}(3,\R)$ is
\[
\sigma_{(x,F)}(A)=(0,FA).
\]
With $\rho(P)=J_P^{-1}=(\det P)^{-1}$ we have $\vartheta(A)=-\tr A$ and
\[
\bpsi\big((x,F),P\big)=W_0(x,F;P)=J_P^{-1}\,W(x,FP),
\]
which is the expression introduced in \cite{EM90} for a uniform body with
implant $P$. The Piola entity is $\Pio=(f,T_0)$, with
\[
f=\frac{\partial W_0}{\partial x}=J_P^{-1}\frac{\partial W}{\partial x}(x,FP),
\qquad
T_0=\frac{\partial W_0}{\partial F}=J_P^{-1}\,\frac{\partial W}{\partial F}(x,FP)\,P^T,
\]
and $\Pio(\sigma_q(A))=T_0:(FA)=(F^TT_0):A$. Therefore, identifying
$\mathfrak{gl}(3,\R)^*$ with matrices through the pairing $:$,
\begin{equation}\label{eq:Eshclassic}
\Esh=W_0\,I-F^TT_0,
\end{equation}
which is precisely the \emph{Eshelby stress} of the static theory
\cite{Eshelby75,Maugin93}. Since $R_{P*}A=AP$, Theorem~\ref{thm:main} gives
$\frac{\partial W_0}{\partial P}:(AP)=-\Esh:A$ for every $A$, that is,
\begin{equation}\label{eq:EMresult}
\frac{\partial W_0}{\partial P}=-\,\Esh\,P^{-T},
\end{equation}
which is the result of Epstein and Maugin \cite{EM90}. For $\rho\equiv1$ (energy
per unit volume of the archetype), the Eshelby entity reduces to $-F^TT_0=
-P^{-T}\big(Z^T\frac{\partial W}{\partial Z}(Z)\big)P^T$, with $Z=FP$, that is, to
minus the \emph{Mandel stress} \cite{Mandel71} of the archetype pulled back to the
reference configuration by the implant; it coincides with minus the Mandel stress
when $P=I$. Finally, since
$\Ad^*_{P^{-1}}\beta=P^{-T}\beta P^{T}$ in matrix form, Corollary~\ref{cor:transf}
reads
\[
\Esh=J_P^{-1}\,P^{-T}\,\Esh_W(FP)\,P^{T},\qquad
\Esh_W(Z)=W(Z)\,I-Z^T\frac{\partial W}{\partial Z}(Z),
\]
the familiar rule for the transformation of the Eshelby stress under a change of
reference configuration.

\subsection{Second-grade bodies}

Let $k=2$ and let the body be ordinary. A $2$-jet $q\in\Q_c$ is represented by
$(x,F,G)$, with $F^i_J=\partial_J\chi^i(c)$ and
$G^i_{JM}=\partial_J\partial_M\chi^i(c)$, symmetric in $J,M$. The group
$\G_c\cong G^2_3$ consists of pairs $(P,Q)$ with $P\in\GL$ and $Q^K_{JM}$
symmetric in $J,M$, and the chain rule gives the composition law
\[
\begin{gathered}
(F,G)(P,Q)=\big(FP,\ G(P\,\cdot,P\,\cdot)+FQ\big),\\
\big(G(P\,\cdot,P\,\cdot)+FQ\big)^i_{JM}=G^i_{KL}P^K_JP^L_M+F^i_KQ^K_{JM}.
\end{gathered}
\]
The Lie algebra $\g_c$ consists of pairs $(A,\Gamma)$, which are the first and
second derivatives at $c$ of a vector field $\xi$ vanishing at $c$:
$A^K_J=\partial_J\xi^K(c)$, $\Gamma^K_{JM}=\partial_J\partial_M\xi^K(c)$.
Differentiating twice the components $F^i_K(X)\,\xi^K(X)$ of $T\chi\circ\xi$ at
$c$, Lemma~\ref{lem:sigmajets} yields
\begin{equation}\label{eq:sigma2}
\sigma_q(A,\Gamma)=\Big(0,\ F^i_KA^K_J,\ G^i_{KM}A^K_J+G^i_{JK}A^K_M
+F^i_K\Gamma^K_{JM}\Big).
\end{equation}
The character \eqref{eq:rho} is $\rho(P,Q)=(\det P)^{-1}$, so that
$\vartheta(A,\Gamma)=-\tr A$. The Piola entity of
$\bpsi(q,(P,Q))=\rho(P,Q)\,\psi(q(P,Q))$ is $\Pio=(f,T,H)$, where $T_i{}^J$ is the
Piola stress and $H_i{}^{JM}$, symmetric in $J,M$, the Piola hyperstress, namely
\[
\Pio(\delta x,\delta F,\delta G)=f_i\,\delta x^i+T_i{}^J\,\delta F^i_J
+H_i{}^{JM}\,\delta G^i_{JM}.
\]
Pairing with \eqref{eq:sigma2} and using the symmetry of $H$ and $G$,
\[
\Pio\big(\sigma_q(A,\Gamma)\big)=
\big(F^i_KT_i{}^J+2\,G^i_{KM}H_i{}^{JM}\big)A^K_J
+F^i_KH_i{}^{JM}\,\Gamma^K_{JM}.
\]
Hence the Eshelby entity $\Esh=(\Esh_K{}^J,\Esh_K{}^{JM})$, acting as
$\Esh(A,\Gamma)=\Esh_K{}^JA^K_J+\Esh_K{}^{JM}\Gamma^K_{JM}$, is
\begin{equation}\label{eq:Esh2}
\Esh_K{}^J=\bpsi\,\delta_K^J-F^i_K\,T_i{}^J-2\,G^i_{KM}\,H_i{}^{JM},
\qquad
\Esh_K{}^{JM}=-F^i_K\,H_i{}^{JM}.
\end{equation}
The first component is the Eshelby stress of a second-grade material, in which
the hyperstress contributes through the second deformation gradient; the second
component is a \emph{hyper-Eshelby} entity, which is the driving force conjugate
to the second-order part of the material evolution. By Theorem~\ref{thm:main},
the derivatives of $\bpsi$ with respect to $(P,Q)$ are obtained from
\eqref{eq:Esh2} through the right translation of $G^2_3$:
\[
R_{(P,Q)*}(A,\Gamma)=\frac{d}{dt}\Big|_{t=0}\exp\big(t(A,\Gamma)\big)(P,Q)
=\big(AP,\ \Gamma(P\,\cdot,P\,\cdot)+AQ\big),
\]
so that, for every $(A,\Gamma)$,
\[
\frac{\partial\bpsi}{\partial P}:(AP)+\frac{\partial\bpsi}{\partial Q}\,\vdots\,
\big(\Gamma(P\,\cdot,P\,\cdot)+AQ\big)=-\Esh_K{}^JA^K_J-\Esh_K{}^{JM}\Gamma^K_{JM}.
\]
Taking $A=0$ we read off
$\big(\partial\bpsi/\partial Q\big)_K{}^{LN}P^J_LP^M_N=-\Esh_K{}^{JM}$, and then,
taking $\Gamma=0$, the derivative with respect to $P$. This is the explicit form
of a computation which becomes rather confusing if attempted directly in
components.

\subsection{Bodies with microstructure}

Let the body be a product bundle $\mathcal C_0=\mathcal B_0\times\R^n$, where
$\R^n$ describes $n$ microstructural parameters (directors, order parameters,
etc.), with physical space $\mathcal S=\E\times\R^n$, and let $k=1$. A
configuration is a fibred map $(X,Y)\mapsto(\chi(X),\Phi(X,Y))$, and its
$1$-jet at $c=(X_0,Y_0)$ is
\[
q=(x,y,F,N,L),\qquad x=\chi(X_0),\ y=\Phi(c),\ F=D\chi(X_0),\
N=\partial_X\Phi(c),\ L=\partial_Y\Phi(c),
\]
with $F\in\GL$, $N\in\R^{n\times3}$ and $L\in GL(n,\R)$. The vertical group
$\G_c$ consists of the triples $g=(A,B,C)\in\GL\times\R^{n\times3}\times GL(n,\R)$
of $1$-jets of fibred maps fixing $c$, and the chain rule gives the
action
\[
(x,y,F,N,L)(A,B,C)=(x,y,FA,\ NA+LB,\ LC).
\]
The Lie algebra consists of triples $(\alpha,\beta,\gamma)$ and
\[
\sigma_q(\alpha,\beta,\gamma)=(0,0,F\alpha,\ N\alpha+L\beta,\ L\gamma).
\]
Measuring energies per unit volume of the base, $\rho(A,B,C)=(\det A)^{-1}$ and
$\vartheta(\alpha,\beta,\gamma)=-\tr\alpha$. Writing the Piola entity as
$\Pio=(f,\varphi,T,S,\Sigma)$, with $T=\partial\bpsi/\partial F$ (the Piola
stress), $S=\partial\bpsi/\partial N$ (the micro-stress conjugate to the
gradient of the microstructure) and $\Sigma=\partial\bpsi/\partial L$, we get
\[
\Pio(\sigma_q(\alpha,\beta,\gamma))=T:(F\alpha)+S:(N\alpha+L\beta)+\Sigma:(L\gamma).
\]
Therefore the Eshelby entity has three components,
\begin{equation}\label{eq:EshMicro}
\Esh=\Big(\bpsi\,I-F^TT-N^TS,\ \ -L^TS,\ \ -L^T\Sigma\Big),
\end{equation}
conjugate, respectively, to the evolution of the macro-implant, to the
evolution of the coupling between macro- and microstructure, and to the
evolution of the micro-implant. The first component is the Eshelby stress of a
generalized Cosserat medium (compare with \cite{Maugin93,EdL98}); the
term $-N^TS$ is the contribution of the gradient of the microstructure. Again,
the material derivatives of $\bpsi$ follow from Theorem~\ref{thm:main} and the
right translation of $\G_c$,
$R_{(A,B,C)*}(\alpha,\beta,\gamma)=(\alpha A,\ \beta A+\gamma B,\ \gamma C)$.

\begin{remark}
For Cosserat media, and more generally for media whose microstructure is
modelled by a principal bundle (for instance, the linear frame bundle of the
body) \cite{Capriz89,EdL98}, configurations are required to be principal bundle
morphisms. The manifold $\Q_c$ and the group $\G_c$ are then replaced by the
corresponding sets of $k$-jets at $c$ of principal bundle morphisms and of
principal bundle automorphisms fixing $c$. The latter is a closed subgroup of the
vertical group used above, it acts freely on the former, and composition
preserves equivariance. Theorem~\ref{thm:main} therefore applies without
change, with $\sigma_q$ restricted to the Lie algebra of this subgroup; if the
energy is the restriction of an energy defined on all fibred jets, the Eshelby
entity is the restriction of \eqref{eq:EshMicro} to that subalgebra.
\end{remark}

\section{Material symmetry and material evolution}\label{sec:sym}

\subsection{Material symmetries}

A closed subgroup $\mathcal H\subset\G_c$ is a \emph{symmetry group} of the
archetype if $\psi(qh)=\psi(q)$ for all $q\in\Q_c$ and $h\in\mathcal H$. We
shall assume that symmetries preserve volume, $\rho|_{\mathcal H}=1$, as is the
case for the usual symmetry groups of solids, so that $\vartheta$ vanishes on the
Lie algebra $\mathfrak h$ of $\mathcal H$. Since
$\bpsi(qghg^{-1},g)=\rho(g)\psi(qgh)=\bpsi(q,g)$, the symmetry group of the
evolving material in the reference configuration is the conjugate group
$g\mathcal Hg^{-1}$, whose Lie algebra is $\Ad_g\mathfrak h$.

\begin{proposition}\label{prop:sym}
The Eshelby entity annihilates the Lie algebra of the symmetry group of the
material in the reference configuration:
\[
\Esh(\Ad_gA)=0\qquad\text{for all }A\in\mathfrak h.
\]
Equivalently, the material entity annihilates the tangent space at $g$ of
the coset $g\mathcal H$, which consists of the directions of material evolution
that leave the response unchanged.
\end{proposition}

\begin{proof}
By \eqref{eq:Pio} and \eqref{eq:Requiv},
\[
\Pio(\sigma_q(\Ad_gA))=\rho(g)\,\psi_*|_{qg}(\sigma_{qg}(A))=
\rho(g)\frac{d}{dt}\Big|_{t=0}\psi(qg\exp(tA))=0,
\]
since $\exp(tA)\in\mathcal H$. Moreover, $\vartheta(\Ad_gA)=\vartheta(A)=0$. For the
second statement, $T_g(g\mathcal H)=L_{g*}\mathfrak h$ and, by
Theorem~\ref{thm:main}, $\Mat(L_{g*}A)=-\Esh(\Ad_gA)=0$.
\end{proof}

For an isotropic simple solid, $\mathcal H=SO(3)$ (conjugated by $g$), and
Proposition~\ref{prop:sym} states that the Eshelby stress
\eqref{eq:Eshclassic} is symmetric with respect to the metric induced by the
implant, which is the well-known symmetry of the Mandel stress for isotropic
materials. For second-grade materials and bodies with microstructure,
Proposition~\ref{prop:sym} provides, without any further computation, the
corresponding restrictions on the components \eqref{eq:Esh2} and
\eqref{eq:EshMicro}.

\subsection{Dissipation and evolution laws}

Let us now consider a process $t\mapsto(q(t),g(t))$ at a fixed material point
$c$, in which both the local deformation and the local material deformation
evolve in time. In an isothermal setting, and neglecting fluxes and sources of
energy and of mass, the local dissipation is the difference between the
internal power $\Pio(\dot q)$, which for $k\ge2$ includes the power of the
hyperstresses, and the rate of change of the stored energy:
\[
\mathscr D=\Pio(\dot q)-\frac{d}{dt}\bpsi(q,g)=-\Mat(\dot g).
\]
Writing $\dot g=R_{g*}\Lambda$, where
$\Lambda=R_{g^{-1}*}\dot g\in\g_c$ is the right-trivialized rate of material
evolution ($\Lambda=\dot P P^{-1}$ in the simple case), Theorem~\ref{thm:main}
gives
\begin{equation}\label{eq:diss}
\mathscr D=\Esh(\Lambda)\ \ge\ 0 .
\end{equation}
Therefore the Eshelby entity is the driving force of material evolution, at any
order and for bodies with or without microstructure. Any evolution law of the
form
\begin{equation}\label{eq:evol}
R_{g^{-1}*}\,\dot g=\Lambda(\Esh),\qquad \Esh\big(\Lambda(\Esh)\big)\ge0,
\end{equation}
is consistent with the dissipation inequality. A simple choice is obtained by
fixing an inner product on $\g_c$ (a mobility tensor) and setting
$\Lambda=\mu\,\Esh\sh$, with $\mu\ge0$ and $\Esh\sh$ the vector associated
with $\Esh$ by the inner product. In the simple case, with the Frobenius inner
product and the identification of $\mathfrak{gl}(3,\R)^*$ with matrices of
Section~\ref{sec:examples}, $\Esh\sh=\Esh$, and this yields
$\dot PP^{-1}=\mu\,\Esh$, for which $\mathscr D=\mu\,\Esh:\Esh\ge0$; this is the type of law proposed in
\cite{EM00,EEbook} for remodelling and growth. By
Proposition~\ref{prop:sym}, the directions $\Ad_g\mathfrak h$ produce no
dissipation, in agreement with the fact that they do not change the material
response.

\begin{remark}
In processes of volumetric growth, additional terms related to the flux and
the source of mass appear in the balance of energy and in the dissipation
inequality \cite{EM00}. They do not affect the identification of the Eshelby
entity as the conjugate of $\Lambda$ provided in \eqref{eq:diss}.
\end{remark}

\section{Concluding remarks}

We have shown that the relation between the material entity and the Piola
entity in theories of material evolution is a direct consequence of the
differential of a free right action of a Lie group, namely the group $\G_c$ of
$k$-jets of local material deformations acting on the manifold $\Q_c$ of
$k$-jets of deformations. The resulting general Eshelby entity
\[
\Esh=-\bpsi\,\vartheta-\sigma_q^*\Pio\in\g_c^*
\]
lives naturally in the dual of the Lie algebra of the vertical group, it
recovers the classical Eshelby stress for simple bodies, and it is completely
explicit for higher-grade materials and for bodies with microstructure, since
the infinitesimal action $\sigma_q$ amounts to composing the jet of the
deformation with jets of vector fields vanishing at the point.

The construction is pointwise. The natural next step is to let $c$ vary, so
that $\Q_c$ and $\G_c$ become, respectively, a principal bundle and the
gauge (isotropy) group bundle of the Lie groupoid of $k$-jets of local
deformations, and the material entities become sections of the dual of the
corresponding Lie algebroid. This would place the present results within the
framework of material uniformity and homogeneity in terms of Lie groupoids and
Lie algebroids developed in \cite{JdLE19,JdLE20,EdL16,dLEJ21,JdLE22}, and should lead to a
geometric formulation of the balance of configurational forces at any order.

\section*{Acknowledgements}
M.~de Le\'on acknowledges financial support from the Spanish Ministry of Science,
Innovation and Universities (MICIU), project PID2022-137909NB-C21, and from the
Severo Ochoa Programme for Centres of Excellence in R\&D (CEX2023-001347-S). He
also acknowledges partial support from the National Natural Science Foundation of
China (NSFC, Grant No.~W2541005), from the Guangdong Provincial Key Laboratory of
Mathematical and Neural Dynamical Systems (Grant No.~2024B1212010004), and from a
Distinguished Visiting Professorship at Great Bay University.

\end{document}